\documentclass[prl,aps,twocolumn,showpacs]{revtex4}
\usepackage{epsfig}
\usepackage{bm} 
\newcommand{\B}[1]{{\bm{#1}}}

\newcommand{\Onecol} {\begin{widetext} \onecolumngrid} 
\newcommand{\Twocol} {\end{widetext} \twocolumngrid} 

\newcommand{\be}{\begin{equation}}
\newcommand{\ba}{\begin{array}}
\newcommand{\bea}{\begin{eqnarray}}
\newcommand{\bfi}{\begin{figure}}
\newcommand{\ee}{\end{equation}}
\newcommand{\ea}{\end{array}}
\newcommand{\eea}{\end{eqnarray}}
\newcommand{\efi}{\end{figure}}

\begin{document} 
\title{On the Fractal Dimension of the Visible Universe}
\author{Jean-Pierre Eckmann$^1$, Esa J\"arvenp\"a\"a$^2$, Maarit J\"arvenp\"a\"a$^2$,
and Itamar Procaccia$^3$}
\affiliation{$^1$Dept.~de Physique Th\'eorique et Section de Math\'ematiques, Universit\'e de
Gen\`eve, Switzerland\\$^2$Dept. of Mathematics and Statistics,  P.O. Box 35 (MaD),
40014 University of Jyv\"askyl\"a, Finland\\$^3$Dept.~of Chemical Physics, The Weizmann
Institute of Science, Rehovot, 76100 Israel} 
\begin{abstract} 
Estimates of the fractal dimension $D$ of the set of galaxies in the
universe, based on ever improving data sets, tend to settle on $D\approx 2$.
This result raised a raging debate due to its glaring contradiction
with astrophysical models that expect a homogeneous universe. A recent
mathematical
result indicates that there is no contradiction, since measurements of the
dimension of the {\em visible} subset of galaxies is bounded from above by
$D=2$ even if the true dimension is anything between $D=2$ and $D=3$. We
demonstrate this result in the context of a simple fractal model, and
explain
how to proceed in order to find a better estimate of the true dimension
of the set of galaxies.
\end{abstract} 
\maketitle
The value of the (fractal) dimension $D$ of the galaxy distribution in the universe
is an important open question in cosmology. Steadily improving 
observations are available, giving scientists 
hope that enough data will allow finally to decide the highly debated issue of
whether $D$ is 3 or substantially lower (usually stated to be about 2).
For example, in the recent book
\cite{02BT} on the ``Discovery of Cosmic Fractals"
it is emphatically declared
that ``The
megafractals -- the cosmic continents, archipelagos and islands -- were the news brought home
by the modern explorers of the cosmos, exotic, but truths
nevertheless about
the worlds overseas. Even if the fractal dimension and the maximum scale are
still debated, megafractals cry for explanation. Their origin is the
number one 
challenge for cosmological physics.'' What these
authors refer to are mainly results of fractal analysis of the data sets of
galaxies
which indicate the fractal dimension $D$ of the set of galaxies
is about 2 \cite{92CP,98SMP,02PBMS}.
In this Letter we build on a recent theorem of fractal
mathematics \cite{02JJMO} which
indicates that these results may not be ``truths nevertheless", but rather
a reflection of an inherent impossibility to measure a dimension larger than
2. The true dimension may be anything between 2 and 3, and the upper number
is not excluded, in agreement with standard astrophysical theories of a
homogeneous universe \cite{Pee}. In this way, ``the
number one challenge" may have been resolved in an unexpected and
somewhat disappointing way,
namely: when $D>2$
{\em one cannot measure} $D$ by observing the visible galaxies.
In addition to explaining this result, we present a partial
remedy by exploring certain aspects of the data analysis that may indicate
the existence of a dimension larger than 2.

In essence the fractal analysis of any given atlas of galaxies is
a simple matter, once one takes carefully into account the side issues
described in \cite{98SMP} which deal with
questions like limited angles, faint luminosities, and other observational
issues. After
worrying about all these details one ends up with a set of points, or coordinates,
each of which stands for a galaxy, with redshift data used to determine
its distance from us (the observers). Given such a set of points $\{\B
X_i\}_{i=1}^N$
in ${\cal R}^3$,
we define the correlation integral $C(r)$ as the number of pairs of points of this
set whose distance is smaller or equal to $r$,
\begin{equation}
C(r) =\frac{2}{N(N-1)}\sum_{i<j} \theta(r-|\B X_i-\B X_j|) \ , \label{corint}
\end{equation}
where $\theta(y)$ is the step function, unity for $y>0$ and zero otherwise.
For a fractal set of dimension $D$ plotting $\log(C(r))$ versus $\log r$
results in a curve whose slope
is the correlation dimension $D_2$ of the Grassberger-Procaccia algorithm
\cite{83GP} (see also \cite{83HP,ER}). In general $D_2\le D$; for sets whose
clustering
is not singular one can expect that $D_2=D$ \cite{83HP}.
For reasons related to angular restrictions and the like, in \cite{92CP,98SMP,02PBMS}
the authors consider a quantity $\Gamma^*(r)$,
which for a general fractal coincides with $C(r)/r^3$. 
Thus their plots should have slopes $D_2-3$. They find consistently
$D_2\approx 2$.

The question is then whether this is really an indication that the set
of all galaxies is of dimension $D\approx 2$. We argue first that this
may not be the case. In a recent paper \cite{02JJMO} the following theorem
was established:  let $F$ be a fractal set in $R^3$ with dimension $D>2$.
The {\em visible part} of the set $F$ from a point $P$ is the subset
$F_V$ of those points lit by a spotlight at $P$. 
{\em Then the part
$F_V$ that is visible to an observer can in general not have a dimension
more than 2} \cite{details}. 

We stress that this
result is not about the projection of the set onto the celestial sphere, but
about those
observations in which the {\em distance} of each point (galaxy) is
given along with its
celestial coordinates (such data sets are called 3-dimensional
catalogs). The meaning of the theorem is that it is in fact impossible to
determine the
dimension of the set of galaxies from measurements of the visible subset if
the
dimension of the full set is larger than 2. The basic reason for this
impossibility is
that galaxies ``hide'' behind each other when the dimension is above 2. This
issue will not go away with improving the catalogs. Rather, it will become
more
and more important as better and better catalogs become available.

We now illustrate some aspects of this problem, and in
particular show that there might be some lower bound on the true dimension
when taking into account finite size effects (which are absent
in mathematical treatments of fractals, but are an evident necessity
of any real-life experiment).

We first note that the catalogs provide measurement of the
positions of galaxies away from us. In other words, we should
consider a relatively small sphere around $P$ and look with radial
rays issuing from the
sphere. In \cite{02JJMO} it is shown that looking from a plane defines an
equivalent problem, and we prefer that formulation. To further
simplify the discussion in our examples we will consider a fractal embedded
in 2 rather than 3 dimensions, illuminated by rays perpendicular to a
randomly given 
baseline.
\begin{figure} 
\includegraphics[width=.4\textwidth]{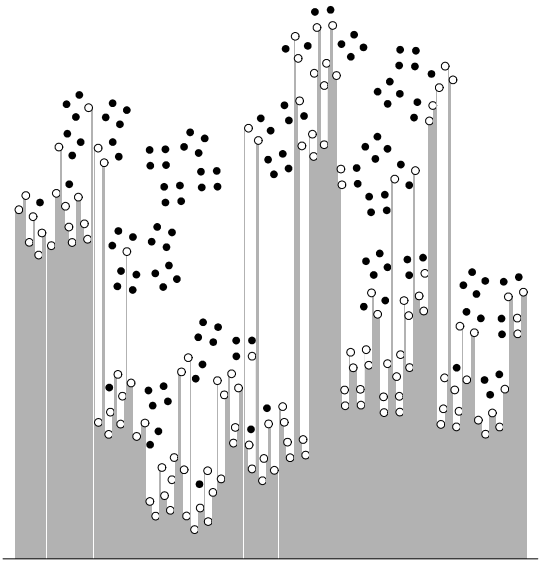}
\caption{A simple illustrative model for a fractal universe, drawn by a
hierarchical construction at level 4, with the visible part in white, the
invisible
points in black. A disk is deemed visible as soon as any part of it is
visible.
Since the construction involves division by 4 and scale changes by
$\lambda =0.4$ the dimension is $D=-\log 4/\log\lambda\approx 1.51 $.}
\label{fig1} 
\end{figure} 
In Fig.~1 we present a simple model of a fractal universe which is
constructed
hierarchically. At the $n$th level of the construction we see $4^n$ balls of
size $\lambda^n$ which are supposed to contain galaxies. At the $(n+1)$th
level each ball is further subdivided to 4 balls of size $\lambda^{n+1}$.
To avoid
non-generic effects we rotate the new group
of balls with a random angle  at each step of construction.
Fig.~1 shows the set of balls at the 4th level with $\lambda=0.4$. The fractal
dimension of this example is $D=-\log(4)/\log(\lambda)=1.5129$. The figure represents the visible
set $F_V$ (from a random line) as the lighted balls, whereas the invisible
set (the complement of $F_V$) is shown as black balls.
One can understand the theorem of \cite{02JJMO} in the following intuitive
sense. The projection (the footprint of the gray zone of Fig.~1)
of the fractal on the line has dimension 1 when
$D>1$ \cite{75Mat,85Fal}. On the other hand, for the hierarchical
construction up to level $k$, the balls have size $\lambda ^k$, and
thus the projection of the visible part has dimension 1 as soon as
there are at least $1/\lambda ^k$ visible balls (assuming they hide
all others). The boundary between the gray zone and the white zone
in Fig.1 forms a graph of a function connecting the visible balls. In the $k$th
level we call this function $f_k(x)$. Since the fractal set is constructed
hierarchically, we expect a scaling relation 
$f_{k+1}(x)=\lambda^{-1} f_k(\lambda ^{-1}x)$. This scaling relation guarantees
that the graph cannot become too rough and will remain of dimension 1.
To see this clearly think about the Weierstrass function
$g(x)=\sum_{k=0}^\infty  a^k \sin(b^k x)$. It is well known that
the graph of this function is rough when $ab>1$ and $a<1$. Indeed,
(cf. \cite{98H}) the Weierstrass function almost scales in the sense that
\begin{equation}
g(x)=a^{-1}g(bx) -a^{-1} \sin x \ . 
\end{equation}
The last term is smooth and its contribution to the dimension is
negligible. Covering the graph with balls leads to the well known
result $D=2-|\log a|/\log b$.
But, in our analogous case, $a=\lambda $,
$b=1/\lambda $, leading to a 1-dimensional graph. Loosely speaking the stretching is
not very strong in the
$y$-direction, and the dimension of the graph (and hence
of the visible set)  remains 1. In fact, the same argument explains why
in dimension $D<1$ the visible part of a fractal has the same
dimension as the fractal itself.

\begin{figure} 
\includegraphics[width=.4\textwidth]{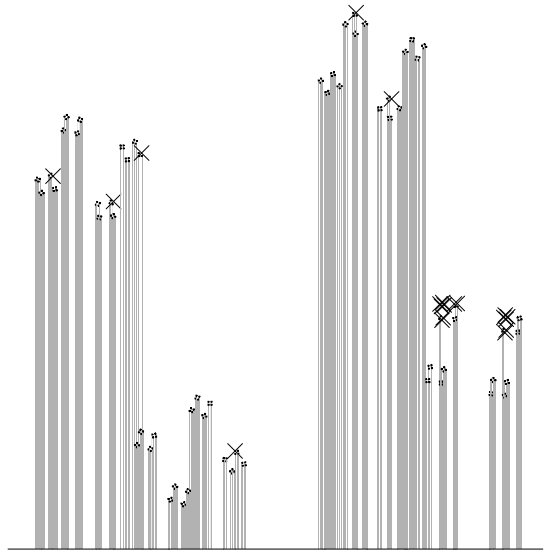}
\caption{A second model, as in Fig.1, but now with dimension
$D=0.85$. Although $D<1$, there are still hidden disks, and
to make them appear more clearly, we marked them with an $\times$.}
\label{fig2} 
\end{figure} 
To illustrate these issues we consider first a fractal of dimension
smaller than 2 (cf.~Fig.~2). Here the visible and full sets will have
the same dimension, as is demonstrated in Fig.~3, where we plot $\log C(r)$
vs. $\log r$ for both $F$ and $F_V$ (upper panel). Evidently the slope is the
same for these sets. To demonstrate this fact further we present in the lower panel
of Fig.~3 a plot of $\log C(r)$ for $F$ vs. $\log C(r)$ for $F_V$.
The slope of this line is unity, stressing the fact that the bulk
of $F$ is revealed in the visible subset $F_V$.
\begin{figure} 
\includegraphics[width=.4\textwidth,angle=270]{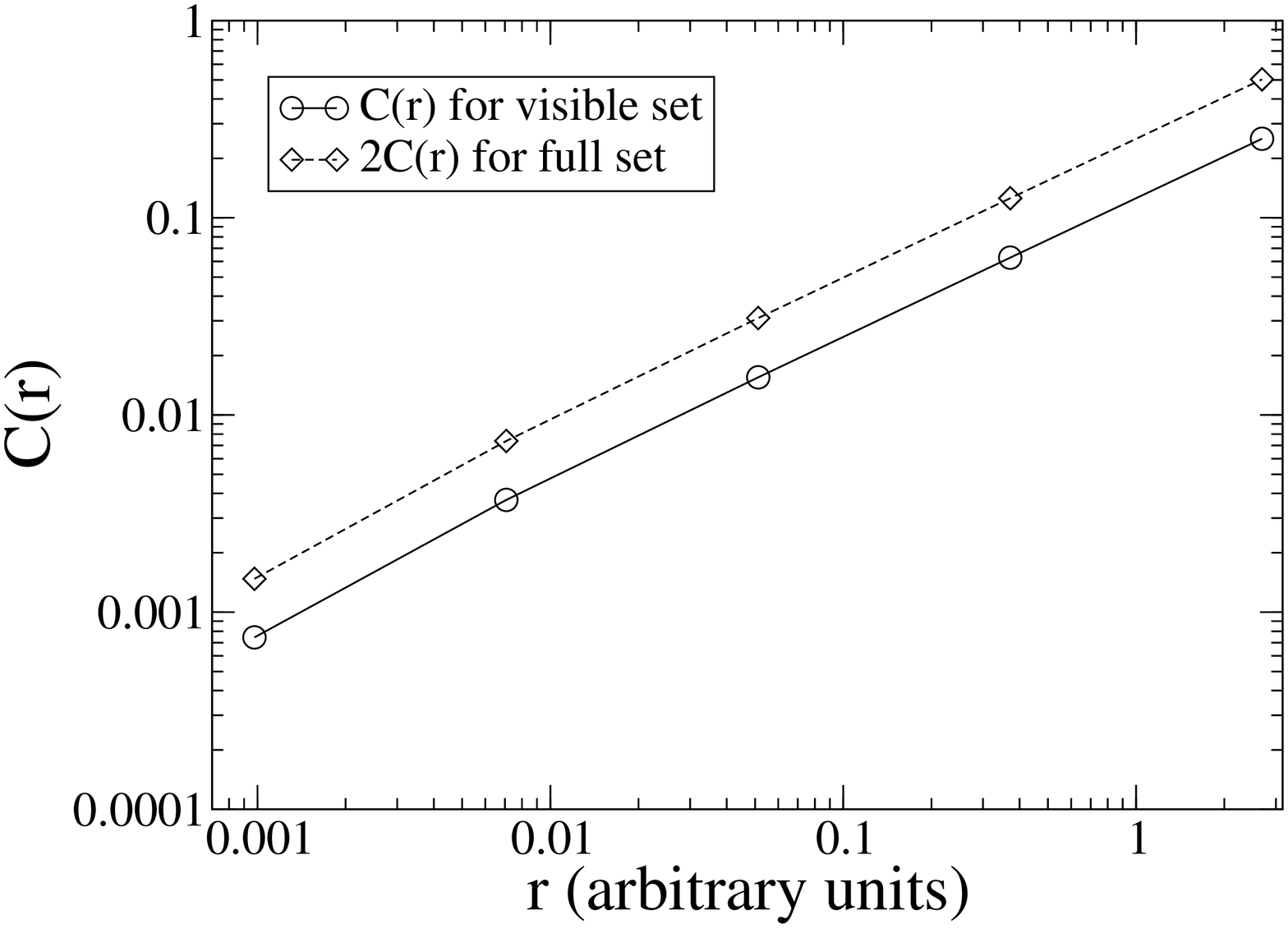}
\includegraphics[width=.4\textwidth,angle=270]{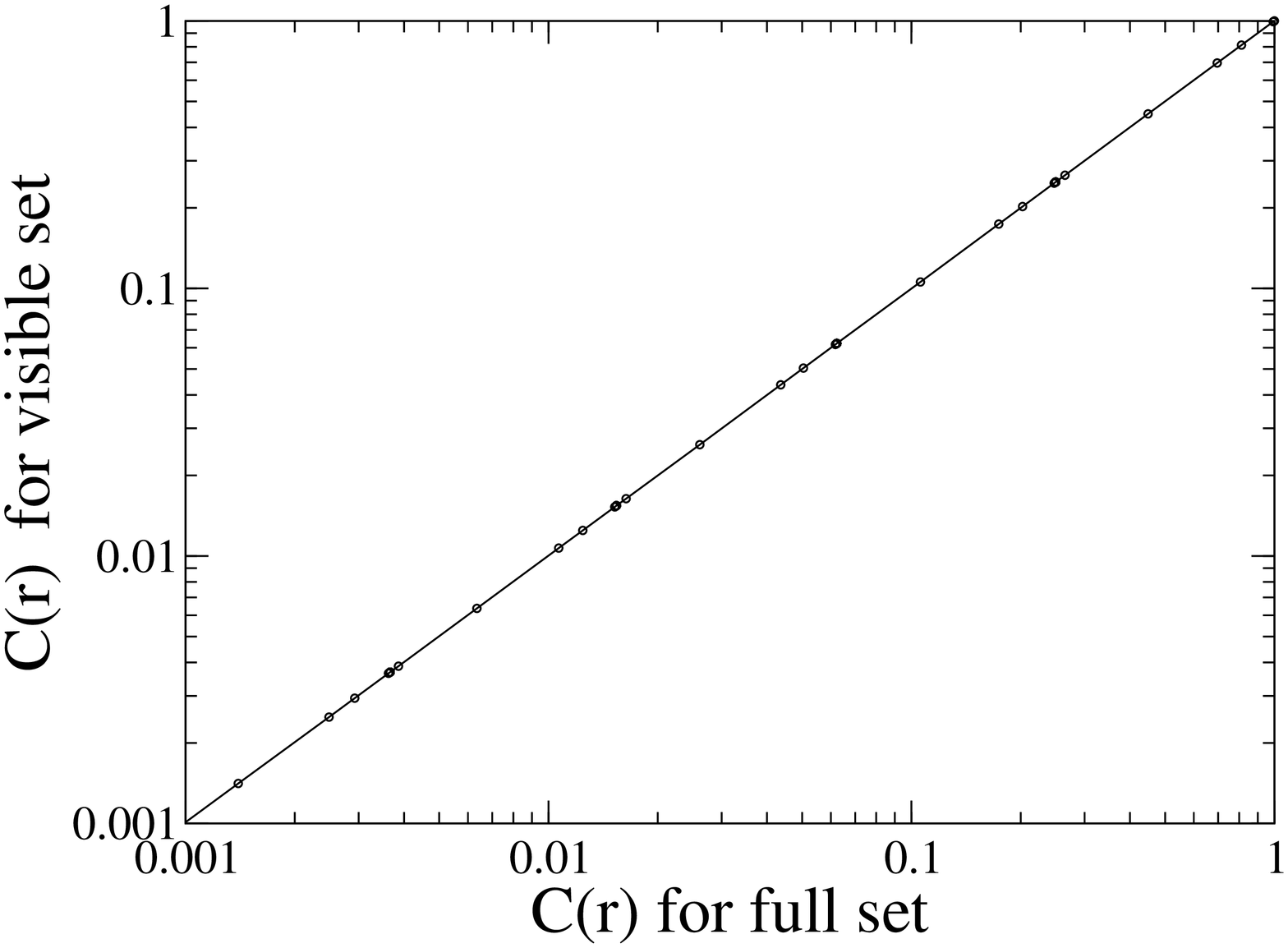}
\caption{Upper panel: The graphs of $C(r)$ for the visible part and the full
fractal of dimension $D=0.7$, at level $6$. The top curve is the
binned number of pairs of points whose pairwise distance falls in the
bin
(in equal bins on the logarithmic scale) for
the full set (multiplied by 2 to shift the curve up). The lower curve
is the same for the visible part. The least square fits for the
measured dimensions are $D=0.7095 \pm 0.0036$ and $D=0.71023 \pm 0.0038$.
Lower panel: $C(r)$ for the full fractal versus $C(r)$ for the visible part
at level $6$. A least square fit gives a slope of $1.0002$. Note that
this does {\em not at all} mean that all disks are visible!}
\label{fig3a} 
\end{figure} 
\begin{figure} 
\includegraphics[width=.4\textwidth,angle=270]{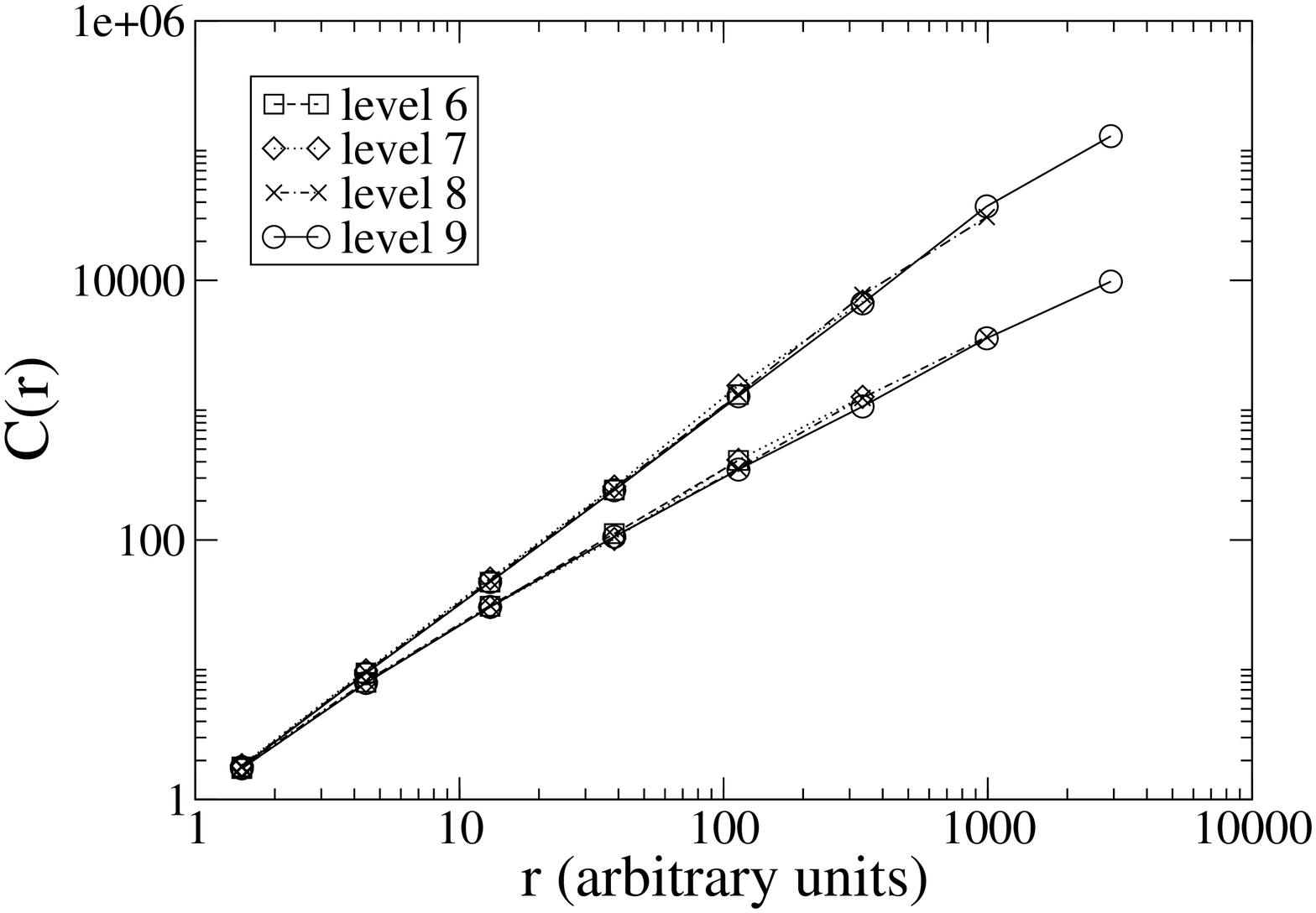}
\includegraphics[width=.4\textwidth,angle=270]{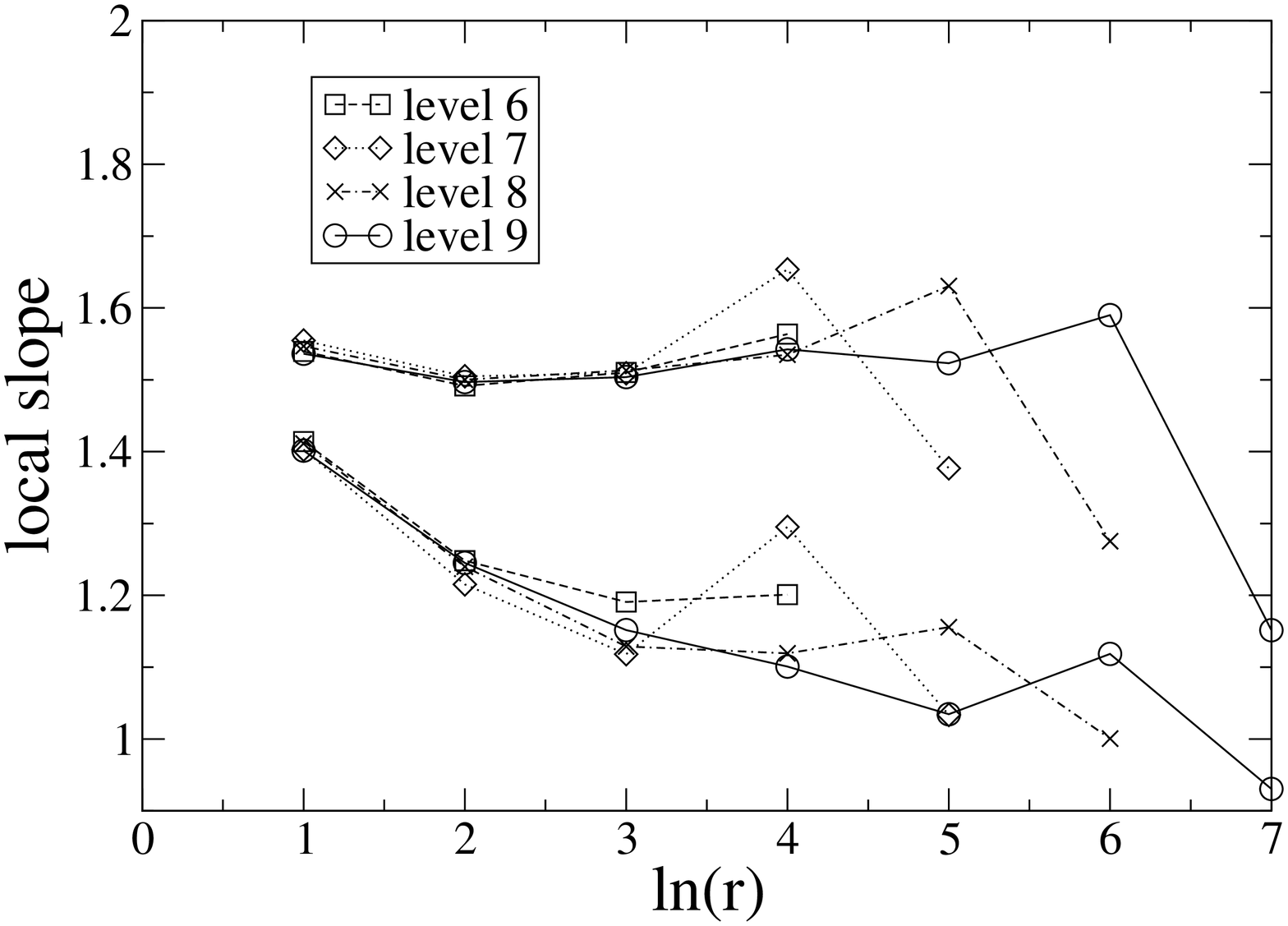}
\caption{Upper panel: Double logarithmic plot of $C(r)$ vs.~$r$ 
for the visible and full sets of dimension $D=1.5$. Shown are
measurements for 4 consecutive levels of hierarchical construction
(sets with $4^6$ to $4^9$ points). The data are normalized
to collapse at the lowest available scale.
Lower panel: The pointwise slopes (dimensions) of the curves in the
upper panel. Note that the full set is seen to have dimension $D=1.5$,
while the visible part tends asymptotically to dimension 1 as the
length scale increases.}
\label{fig4a} 
\end{figure} 

The results change qualitatively when the dimension of the set is higher
than
2. In the upper panel of Fig.~4 we present the double logarithmic plot of the
correlation integral vs distance for $F$ and $F_V$ of the set
of dimension 1.5 of Fig.~1. Clearly they do not scale in the same way,
with the visible set settling on dimension 1 when larger and larger $r$
are taken into account. This is underlined again by the results shown
in the lower panel where the pointwise slopes of the curves in the
upper panel are shown. Obviously the correlation integral for $F$ 
settles nicely on dimension $D\approx 1.5$, whereas the local slope of
the correlation integral for $F_V$ tends to $D\approx 1$ as $r$ increases.
We stress that subdividing the set further in the hierarchic construction
will not cure the problem. Quite on the the contrary, it will make the
visible set $F_V$ a relatively smaller subset of the full set $F$.
Unfortunately going deeper in the hierarchic construction is analogous
to studying larger and deeper catalogs, so we cannot expect that
newer and better data on the galaxy distributions may automatically cure
the problem. We thus conclude that the results of the fractal analysis
presented so far do not exclude a homogeneous universe with the fractal
dimension of the full set of galaxies being as high as 3.

Lastly, we should investigate whether all is lost, or whether there is a way
to probe the true dimension of the full set $F$ from the knowledge of
the visible set $F_V$. A modest way out is offered by the observation that
the slopes of the curves in the upper panel of Fig.~4 {\em are very close}
at {\em small} distances. This observation is underlined by the pointwise slopes 
of the curve in Fig.~4 at small distances. This is clearly a finite size
effect which can be understood by looking again at Fig.~1. Due to the finite
size of the smallest balls at this level of construction, many of
the visible balls appear in groups of 4. This is due to the balls that
were visible for the previous level of construction (4th in this case),
mainly near the edge, but not only, which remain visible also
after one step of refinement. With less degree of conviction one can also
observe groups of 16 balls, or almost 16 balls, that are visible
mainly near the visible edge. This finite size phenomenon will go away
at the present small length scales when we subdivide many steps further,
but will remain observable at the smallest available scales forever. This
observation rationalizes why we get the ``correct" dimension of the
full set $F$ from the smaller scales of the correlation integral.

These observations indicate that despite the mathematical impossibility
of measuring the true
dimension, its value  may be gleaned from the behavior of the
correlation integral 
at small scales. We stress that this possibility is not only due to
close points 
(or galaxies) near the visible edge---also points that are far away
from the observation line (or point $P$) contribute. Balls that are
lighted at level $n$ have high probability to give rise to a full
set of lighted balls also in the next level, but not so for many
subdivisions. Thus lighted balls will
count the ``right" dimension only with regard to small pairwise distances
close to where they are. Once we try to count larger pairwise distances
we unavoidable run to the problems explained above. Indeed, interestingly
enough, it appears that the data analysis presented
in \cite{92CP,98SMP,02PBMS} indicates a slight {\em increase} in the apparent
dimension for smaller scales. We suggest that this increase may very
well point to the true answer, namely, that the dimension of the set
of galaxies is considerably larger than 2, and maybe even 3-dimensional
in agreement with the expectations expressed in \cite{Pee}.

This work was supported in part by the Fonds National Suisse,
the Minerva Foundation, Munich, Germany, the European Commission
under a TMR grant "Control, Synchronization and Characterization of spatially
Extended Nonlinear System" and the Academy of Finland (projects 46208 and 48557).


\begin{thebibliography}{99}
\bibitem{details}In fact, there are really two possibilities of which
the second is even more dramatic for any measurement than the
first:
Either the part $F_V$ that is visible to an observer at $P$ can in general
not have a dimension greater than 2 or it may happen that the dimension of
$F_V$ depends in an essential way on the observation point $P$. All the known 
examples belong to the first class.
\end{thebibliography}
\end{document}